**Title** Versatile Method of Engineering the Band Alignment and the Electron Wavefunction Hybridization of Hybrid Quantum Devices

*Guoan Li, Xiaofan Shi, Ting Lin, Guang Yang, Marco Rossi, Ghada Badawy, Zhiyuan Zhang, Jiayu Shi, Degui Qian, Fang Lu, Lin Gu, An-Qi Wang, Bingbing Tong, Peiling Li, Zhaozheng Lyu, Guangtong Liu, Fanming Qu, Ziwei Dou, Dong Pan, Jianhua Zhao, Qinghua Zhang, Erik P. A. M. Bakkers, Michał P. Nowak,\* Paweł Wójcik,\* Li Lu,\* and Jie Shen\**

G. Li, X. Shi, T. Lin, G. Yang, Z. Zhang, J. Shi, D. Qian, F. Lu, A.-Q. Wang, B. Tong, P. Li, Z. Lyu, G. Liu, F. Qu, Z. Dou, Q. Zhang, L. Lu, J. Shen
Beijing National Laboratory for Condensed Matter Physics, Institute of Physics, Chinese Academy of Sciences, Beijing 100190, China
E-mail: shenjie@iphy.ac.cn; lilu@iphy.ac.cn

M. P. Nowak
AGH University of Krakow, Academic Centre for Materials and Nanotechnology, al. A. Mickiewicza 30, 30-059 Krakow, Poland
E-mail: mpnowak@agh.edu.pl

P. Wójcik
AGH University of Krakow, Faculty of Physics and Applied Computer Science, al. A. Mickiewicza 30, 30-059 Krakow, Poland
E-mail: Pawel.Wojcik@fis.agh.edu.pl

M. Rossi, G. Badawy, E. P. A. M. Bakkers
Department of Applied Physics, Eindhoven University of Technology, 5600 MB Eindhoven, The Netherlands

D. Pan, J. Zhao
State Key Laboratory of Superlattices and Microstructures, Institute of Semiconductors, Chinese Academy of Sciences, P. O. Box 912, Beijing 100083, China

L. Gu

Beijing National Center for Electron Microscopy and Laboratory of Advanced Materials, Department of Materials Science and Engineering, Tsinghua University, Beijing 100084, China

G. Li, X. Shi, T. Lin, Z. Zhang, A.-Q. Wang, F. Qu, L. Lu
School of Physical Sciences, University of Chinese Academy of Sciences, Beijing 100049, China

B. Tong, P. Li, Z. Lyu, G. Liu, F. Qu, Z. Dou, L. Lu, J. Shen
Songshan Lake Materials Laboratory, Dongguan 523808, China

J. Shen
Beijing Academy of Quantum Information Sciences, Beijing 100193, China

D. Pan, J. Zhao
Center of Materials Science and Optoelectronics Engineering, and CAS Center for Excellence in Topological Quantum Computation, University of Chinese Academy of Sciences, Beijing 100190, China

Hybrid devices that combine superconductors (S) and semiconductors (Sm) have attracted great attention due to the integration of the properties of both materials, which relies on the interface details and the resulting coupling strength and wavefunction hybridization. However until now, none of the experiments have reported good control of band alignment of the interface, as well as its tunability to the coupling and hybridization. Here, the interface is modified by inducing specific argon milling while maintaining its high-quality, e.g., atomic connection, which results in a large induced superconducting gap and ballistic transport. By comparing with Schrödinger-Poisson calculations, it is proven that this method can vary the band bending/coupling strength and the electronic spatial distribution. In the strong coupling regime, the coexistence and tunability of crossed Andreev reflection and elastic co-tunnelling—key ingredients for the Kitaev chain—are confirmed. This method is also generic for other materials and achieves a hard and huge superconducting gap in Pb-InSb devices.

Such a versatile method, compatible with the standard fabrication process and accompanied by the well-controlled modification of the interface, will definitely boost the creation of more sophisticated hybrid devices for exploring physics in solid-state systems.

**1. Introduction**

The metal-semiconductor hetero-interface critically determines the performance of modern electronic devices. When a semiconductor and a metal contact, an interface where energy levels are discontinuous is formed. Its details result in different charge densities and electric field distribution crossing the semiconductor and metal, which control the charge current and response through the entire device. That is, the band alignment at the metal-semiconductor interface greatly influences the nature of the contact, with accumulation band alignment leading to ohmic contact favorable for device operation, while depletion of the interface creates a Schottky barrier that poses significant challenges.[1] Therefore, optimization of the hetero-interface is of great significance to improve the electric properties of such hybrid devices.

In particular, the advancement of quantum technology and topological physics have sparked considerable interest in hybrid devices of superconductors and semiconducting nanowires with spin-orbit coupling owing to their potential realization of Majorana zero modes (MZMs) and topological quantum computation. The nonlocal nature with topological protection for MZMs holds promise in robustly storing and manipulating quantum information, making them a sought-after candidate for fault-tolerant topological qubits.[2-5] Recently, gatemons,[6-7] Andreev spin qubits,[8-11] and other novel hybrid qubits[12-14] also emerge, stemming from advances of the proximitized superconducting condensate with electrostatically gate-tunable and spin-resolved subgap states. In addition, semiconducting quantum dots coupled to superconductor could behave as the single electron filter to separate entangled electrons of Cooper pair and test Einstein-Podolsky-Rosen paradox,[15-16] as well as recently-developed artificial Kitaev chain to realize MZMs.[17-18] Thus, they play an essential role in solid-state physics for the operation of electrons and the realization of advanced quantum devices. In such S-Sm heterostructures, the coupling of the two materials, which relies on the interface band alignment, determines the hybridization of the wavefunctions and the integrated electronic properties including the induced superconducting gap, Landé $g$ factor, and spin-orbit coupling (SOC) strength. So it is of fundamental importance but still lacks systematic studies because of the challenging control of mesoscopic details of the electrostatic potential at the interface and the sensitivity of quantum phenomena to disorder.[19]

Recently, thanks to the motivation for constructing topological qubits, the hybridization at the interface and the detailed corresponding electron distribution of wavefunction in the cross-section when placing a parent superconductor on top of a semiconducting nanowire have been quantitively modeled with Thomas-Fermi approximation and Schrödinger-Poisson (SP) calculation.[20-24] All the crucial physical ingredients of these systems have also been carried out, such as band offset (alignment) between the S-Sm interface, the electrostatic environment, and the orbital effect of the magnetic field. For example, the accumulation layer of the interface is the prerequisite for strong coupling between S-Sm and a large induced gap similar to that of the parent superconductor, which results in desirable high-quality hybrid devices for Majorana qubits.[25-30] However, although the electrostatic environment and orbital effect have already been experimentally used to change the size of the induced gap and effective $g$ factor,[20, 31-33] the band offset is always fixed by the difference of the electron affinity of the semiconductor ($\chi_{SM}$) and the metal work function ($W_M$) and induces the unresolvable disadvantage such as the bad hybridization if it is negative, namely depleted. Consequently, searching for a controllable method to tune the band offset and achieve ideal hybridization is in demand.[20, 23]

On the other hand, in addition to removing the native oxide on the semiconductor due to exposure to air, an ideal smooth and sharp S-Sm interface has been demonstrated to be the basis for creating a hard superconducting induced gap, that is, without any in-gap quasi-particle states which hinder the signal for MZMs and destroy the topological protection.[34-36] The first generation fabrication process involves an ex situ process that consists of etching to remove oxide and subsequent metal deposition, and often results in a soft and small induced superconducting gap.[2] Additionally, the device requires sulfur passivation methods to exhibit ballistic transport.[37-38] Then the second generation has been induced,[20] including either molecular beam epitaxy (MBE) in situ growth[25, 39] or hydrogen cleaning with specific shadow-wall technique[40-41] to achieve atomically-connected interface and hard gap. The bare nanowire segments are typically formed either by chemically etching the superconductors or by using shadowing techniques during superconductor deposition. However, chemical etching can damage the semiconductor crystal and cause chemical contamination.[42] The shadow technique requires a complex design of the mask pattern.[40-41, 43] Furthermore, neither is fully compatible with the state-of-the-art lithography technique (While it is possible to perform top-down fabrication techniques after taking the sample out of MBE, this approach tends to degrade the integrity of the S-Sm interface.). Currently, the configurations of theoretical schemes for relevant quantum computation in semiconductor-superconductor hybrid nanowire systems[3-5] are highly complex and present significant challenges for device fabrication. In the Majorana box qubit scheme,[4] three additional quantum dots and an interference link are required for a single-qubit device to perform full single-qubit control and readout of all Pauli operators, not to mention the scalable designs for topological quantum computation networks.[5] Moreover, the artificial Kitaev chains require an increased number of quantum dots to enhance noise immunity and achieve a longer dephasing time.[44] Hence, with the increasing complexity of novel qubits, it is vital to develop a generic approach to create a smooth interface, tune the band offset, and fabricate advanced structures.

Here, by well-controlling the argon milling parameters, we find the band offset between the aluminum and InSb nanowire (Al-InSb), whose value is naturally negative,[37, 45] could be modified by the milling time, that is, inducing accumulation layer with long milling time. The positively increasing band offset for the accumulation layer results in the different electrotactic environment and the demanded properties of the hybrid devices, such as the highly transparent hetero-interface signified by the large induced gap and nice multiple Andreev reflections (MAR), as well as the ballistic electrical transport with quantized conductance plateaus. Electronic structure calculation from the Schrödinger-Poisson model using the experimental parameters also reveals the variation of wavefunction hybridization with milling time and back-gate voltage as expected. Moreover, in the more advanced hybrid devices with the strong coupling, nonlocal signals further confirm the existence of an induced gap, which enables tunability of crossed Andreev reflection (CAR) and elastic co-tunnelling (ECT)—key ingredients for the realization of Kitaev chain and quantum entanglement probing. Finally, with this protocol, we measure the hard and large gap in the lead (Pb) and InSb devices. In short, it turns out to reveal a versatile method compatible with the state-of-the-art lithography fabrication process, instead of growing in situ by MBE or with the assistance of shadow-wall technology, to achieve the high-quality hybrid system and provide more flexibility towards complex structures including topological qubits.

## 2. Results
### 2.1. Atomic-Resolved Hetero-Interface with Argon Milling

The high-mobility InSb nanowires are grown by metal organic vapor phase epitaxy (MOVPE)[46] and transferred from the growth chip to the silicon substrates with micro-manipulator tip under the optical microscope. The silicon substrates are heavily doped and covered with 300 nm-thick $SiO_2$, which works as the dielectric layer of the back-gate. The patterns of leads are defined with electron beam lithography (EBL). Before evaporation with Al, the oxide layer of the nanowires (usually 1.5-2 nm, see **Figure** 1e) is removed by argon plasma. The milling parameters, such as power, pressure, and accelerating voltage, have been optimized to have a very gentle milling rate (~5 nm/min, the shortest milling time ($t$) to remove all the native oxidation layer is 20 s, see Figure S1) to keep atomically flat surface (Figure 1h,i). Then after the milling process, the chips are directly transferred to the chamber with ultrahigh vacuum to electron beam evaporate Al film with a typical thickness of 120 nm.

The scanning electron microscope (SEM) image of a typical device with $t$ = 40 s is shown in Figure 1a. Figure 1b illustrates a three-dimensional (3D) sketch of the device and the corresponding transport measurement configuration. Based on the Octavio-Tinkham-Blonder-Klapwijk (OTBK) model and the proximity effect,[47] our S-NW-S Josephson junctions (JJ) can be depicted as S/S'/N/S'/S structure (Figure 1c), where S' denotes the proximitized nanowire part (light blue rectangle) induced by the top superconducting electrode, N is the semiconducting region not covered by the superconductor, and the slash represents the interface between the different parts. To facilitate the distinction, we refer to the interface between superconductor and semiconductor nanowires as the hetero-interface (indicated by the black horizontal solid line), while the interface between the covered and uncovered nanowire is called the intrinsic interface (indicated by the white vertical dashed line). Notably, due to the fabrication process (discussed in the following section), the etched facets also exist at both sides of the superconducting electrode, expressed as etched surface (indicated by the black horizontal dashed line).

The cross-section cut of the device prepared by focused ion beam (FIB) is displayed by the aberration-corrected scanning transmission electron microscope annular bright-field (STEM-ABF) image (Figure 1d), and a clear atomic-resolved hetero-interface is revealed on the top and part of the side facets (Figure 1h,i; and Figure S1). Apparently, there is no significant fluctuation and roughness on the facet of the etched InSb nanowire. The energy-dispersive X-ray (EDX) mapping of the device cross-section for different elements reveals that the interface between Al and the nanowire is oxygen-free (Figure 1e). Meanwhile, the bottom side facets still possess the native oxide, indicating that the milling is anisotropic. For clarity, from the composite image of the EDX elemental mapping (the gray box in Figure 1f), we extract the line scan (Figure 1g). The almost zero intensity of the oxygen element indicates the clean Al-InSb interface without oxygen. Furthermore, the abrupt transition in In, Sb, and Al elements intensity at the interface confirms the sharp interface. Importantly, even with long milling time up to 2 min and part of the nanowire has been etched away, this high-quality interface can still be kept (Figure S1). We notice for $t = 20$ s, the hetero-interface still have little oxide left because of not-enough milling time (Figure S1). Due to the smooth and clean surface, ideal superconducting films can be deposited on it. The etched InSb nanowires, the Al film, and the interface between the two are free of visible impurities, defects, and disorder, ensuring strong uniform coupling strength at the Al-InSb interface, resulting in a large induced gap and the ballistic transport, which will be analyzed later.

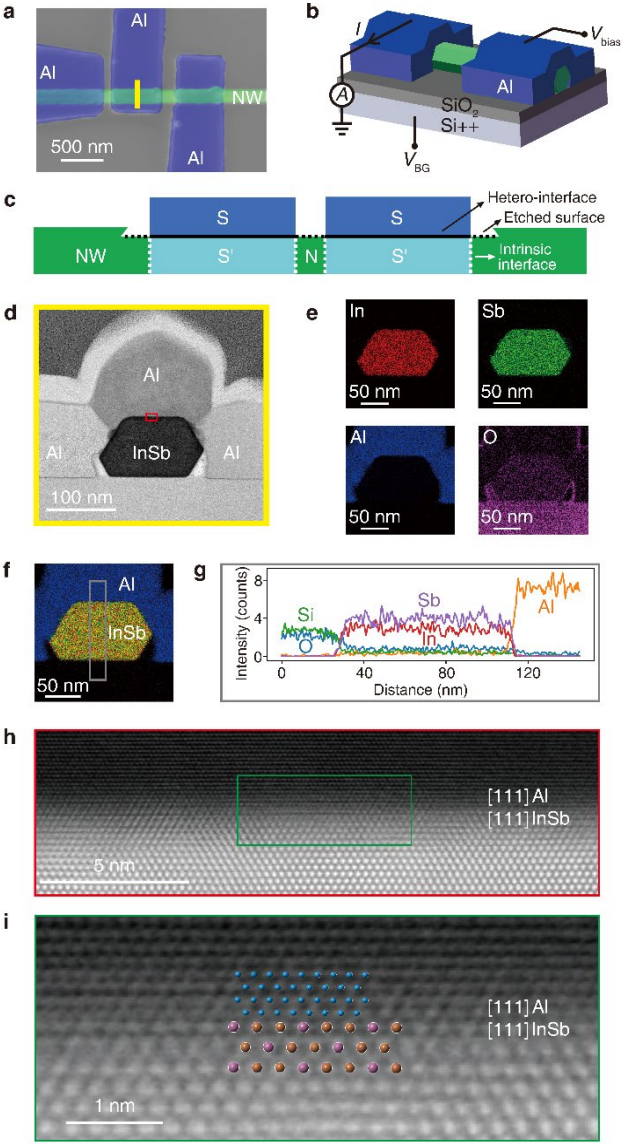

**Figure 1.** Material analysis of a typical device (D2). a) False-color SEM image of the device. Three Al electrodes (blue) contact the InSb nanowire (green), forming a S-NW-S device. b) Schematic of the device and measurement configuration with bias voltage $V_{bias}$, current $I$, and back-gate (Si++ substrate) voltage $V_{BG}$. c) The cross-sectional schematic of the device. S' denotes the proximitized nanowire part (light blue rectangle) induced by the top superconducting electrode. N is the semiconducting region not covered by the superconductor. The black solid lines indicate the hetero-interface between S and S'. The white dashed lines indicate the intrinsic interface between S' and N. The black dashed lines indicate the uncovered etched surface. d) STEM-ABF image of the cross-section, the position is marked with a yellow line in (a). e) Corresponding EDX mapping for different elements. f) The composite image of the EDX elemental mapping. g) Line scan of the integrated elemental counts within the gray box in (f). h,i) STEM-ABF image (contrast inverted) of the Al-InSb interface alone [111] axis at the location indicated by the red and green box in (d,h), respectively. The blue, purple, and brown dots indicate the aluminum (Al), indium (In), and antimony (Sb) atoms, respectively.

## 2.2. The Transparent Interfaces and Ballistic Transport of the Hybrid Devices

We measure the differential conductance as a function of source-drain bias voltage ($V_{bias}$, Y axis) and back-gate voltage ($V_{BG}$, X axis) on the Al-InSb-Al Josephson junctions (JJ) at 10 mK in the dilution refrigerator. Promisingly, the devices with different milling time from 20 s to 2 min (Device 1-3, abbreviated as D1-3 thereafter) always show ballistic transport with quantized above-gap conductance plateaus, as well as nice proximity superconductivity with multiple Andreev reflections and zero-bias conductance peak from supercurrent (**Figure** 2a-c). The boundaries for the first (2Δ', Δ' is the gap induced by the Al film), second (Δ'), and third (2Δ'/3) order of MAR are clearly seen with varying chemical potential (indicated by the red arrows). The line cuts at high $V_{bias}$ in the normal regime above the gap from these three devices are also drawn in Figure 2d,f,h to show the plateaus. We notice the entire 2D map of JJ with $t$ = 20 s still shows some fluctuation, compared to the more smooth maps of the other two with longer $t$, in agreement with the slight residual of oxidate at the interface (Figure S1). Interestingly, the plateaus from JJ with $t$ = 20 s are $G_0$, $3G_0$ and $4G_0$, with the missing of $2G_0$ (the horizontal line cut from Figure 2a shown in Figure 2d), where $G_0 = 2e^2/h$, consistent with the degeneracy of subbands[48] when a nanowire has the hexagonal shape of cross-section. As a contrast, the JJ with longer $t$ show higher plateaus ($4G_0$, $6G_0$, $8G_0$, $10G_0$) and conductance at the similar $V_{BG}$ range, which is a signature of enhanced accumulation band offset (the horizontal line cut from Figure 2c shown in Figure 2h (in blue)). We can also observe the conductance traces at zero bias reproduce the subgap dip structures in different devices (indicated by the purple arrows in Figure 2b,c,f,h). This feature is due to the mixing between the two neighboring subbands and the enhancement of density of states near the opening of the next channel, called van Hove singularity.[37]

To estimate the transparency of the junctions, we fit the MAR traces with numerically obtained curves from a short-junction scattering model (refer to Supporting Information).[49-50] Typical fits of the MAR curves are shown in the right panels of Figure 2a-c. The fitting curves (in black) agree well with the experimental curves (in red). The experimental curves deviate from the fit at smaller $V_{bias}$ (so not shown), probably because the junctions are between the short and long limit. By fitting the MAR curves at different $V_{BG}$, the number of conducting modes and the corresponding transmission coefficients $T$ for three devices can be obtained as a function of $V_{BG}$ (Figure 2e,g,i). As $V_{BG}$ increases, we observe an increase in conductance as a result of the increase in the transmission probability of subsequent modes, which ultimately reaches 1 before the next conduction channel opens. The perfect transmission of the conduction channels signifies the transparent intrinsic interface and gate-tunable normal region. The corresponding fitting above-gap conductance curves are plotted with the black curves in Figure 2d,f,h, which also conform well to the experimental data measured at a high bias. From the fitting, we also obtain the value of the induced gap that determines the position of the MAR structure in the conductance trace (Figure 2a-c). We estimate that the induced gap varies between 0.17 to 0.19 meV (Figure S2), which is close to the gap of parent Al, signifying a good coupling between the nanowire and the superconductor.[35] Moreover, we find that the mean free path (100-200 nm) determined by fitting and calculations is larger than the length of the junctions (~100 nm), further proving the ballistic transport (refer to Supporting Information and Figure S3). As such, by combining such well-developed plateaus and large induced gaps, we conclude that all these hybrid devices display high-quality ballistic transport behaviors and transparent hetero-interfaces.

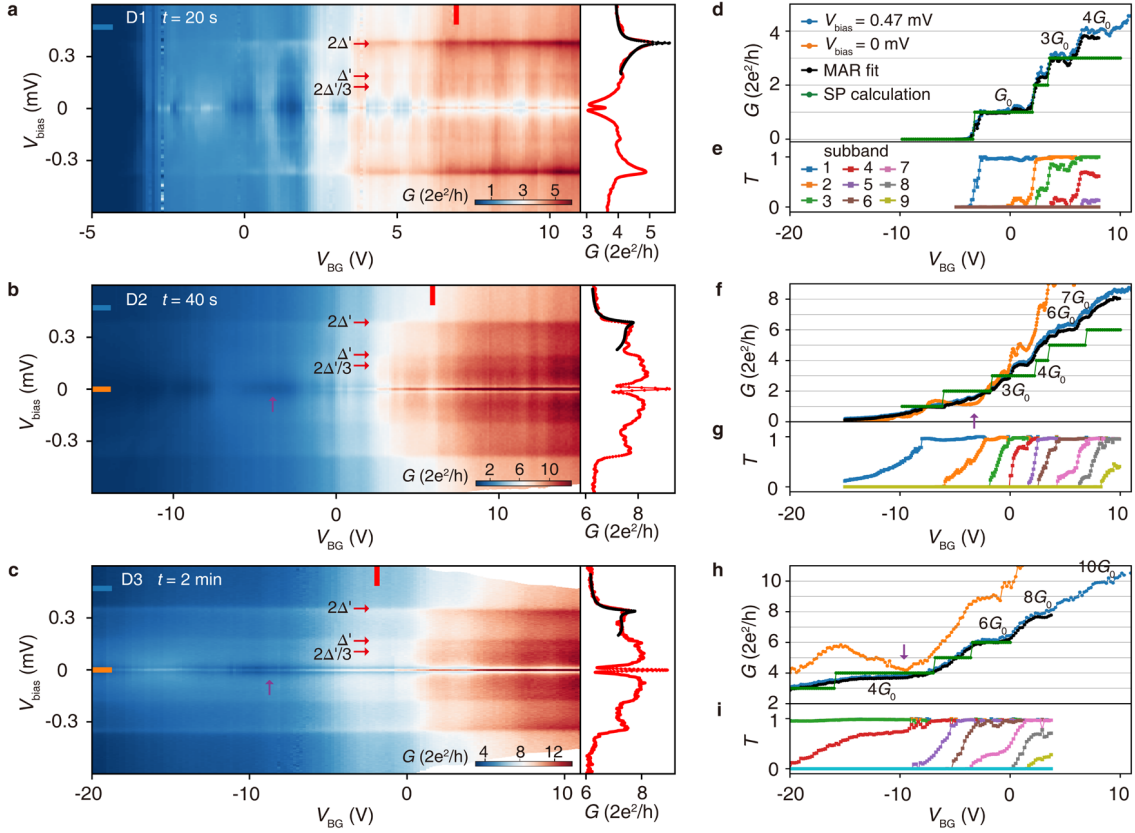

**Figure 2.** Transport measurement of three S-NW-S devices with different milling time $t$ = 20 s, 40 s, 2 min. a-c) Conductance as a function of $V_{bias}$ and $V_{BG}$ for these three devices. The red arrows indicate the boundaries for the first (2Δ'), second (Δ'), and third (2Δ'/3) order of MAR. The purple arrows indicate the subgap dip structures. The red curves on the right panel show typical MAR curves as a function of $V_{bias}$, the vertical line cuts from 2D maps at $V_{BG}$ marked with red bars. The black curves show the fitting of the MAR. d,f,h) Conductance as a function of $V_{BG}$, the horizontal line cuts from panels (a-c) at $V_{bias}$ marked with color bars. To compare with the experiment, the conductance obtained from the MAR fit and Schrödinger-Poisson calculation is shown by the black and green curves, respectively. e,g,i) The transmission of different subbands from MAR fit as a function of $V_{BG}$ in three devices.

## 2.3. Comparison of the Theoretical Model to Reveal the Varying Hybridization of Wavefunction

To give a quantitative description, we also calculate the electronic structures and electron distribution of the normal segment not covered by the Al film with Schrödinger-Poisson method (refer to Supporting Information). In principle, the envelope function approximation (in our case in the form of the SP model) assumes that the envelope of a forward-traveling wave varies slowly in space compared to the wavelength. This condition is satisfied in our study, as the size of our smallest device are comparable to the Fermi length in InSb.[51] The absence of the Al shell in the calculations is because the measured electronic structure is mainly related to the etched bare nanowire segments. The calculation includes electron-electron interaction in the mean field approximation and takes exactly the same parameters as the experiment, i.e., dielectric layer thickness, nanowire width, etc. In particular, the cross-section shape of the nanowires in the calculations varies with the milling time, and the specific thickness of the etched parts is extracted from the STEM-ABF images (Figure 1d; and Figure S1). **Figure** 3a,c,e show the calculated electronic structures as a function of the back-gate voltage of three devices with milling time $t$ = 20 s, 40 s, 2 min. The original Fermi level is determined by the measured conductance at $V_{BG}$ = 0 V in Figure 2a-c. We can note that the electronic structures of the three devices are different, which suggests that the degeneracy of the eigenstates and their positions in the back-gate voltage are closely associated with the shape of the nanowires. Obviously, for longer milling time, the deviation from hexagonal symmetry leads to the removal of the degeneracy of the first and second excited states due to the increased thickness of the etched parts on the top and two upper side facets. Additionally, the electronic structures determine when and which plateaus may appear. For example, when the Fermi level crosses the degenerate points in the electronic structures as the back-gate voltage increases, the value difference between the two neighboring plateaus will be $2G_0$, which has been observed in the transport results in Figure 2d,f,h. For comparison with the experiment measurement and MAR simulations, we plot the conductance plateaus for different milling time with the green curves in Figure 2d,f,h, and find the three are consistent (The slight deviation in conductance between the SP calculations and the experimental data is attributed to the SP calculations not accounting for the specific transmission of the modes during transport). The need for more negative back-gate voltage to pinch off the nanowires at longer milling time reproduced in the simulations should be related to the electronic spatial distribution close to the surface.

Figure 3b,d,f show the electronic spatial distribution obtained by integrating all wavefunctions below the Fermi energy according to the electronic structure at $V_{BG}$ = -8 V, 0 V, and 8 V. At negative $V_{BG}$, the electrons are pushed to the top surface, which would imply a distribution close to the hetero-interface (left panels in Figure 3b,d,f). This convergence of electrons density at the interface between the nanowire and superconductor usually means a strong coupling, which helps to induce a proximitized superconducting gap similar to that of the parent superconductor. Whereas, the positive $V_{BG}$ pulls the electrons toward the bottom facets

near the back-gate. Because long milling time induces accumulation electron density at the etched surface as discussed before, the electrons distribution tends to be like a hollow circle with increasing $V_{BG}$ when $t = 2$ min (middle panel in Figure 3f), compared to a solid circle with more density at the center when $t = 20$ s (middle panel in Figure 3b). As shown in Figure 3a,e, only one subband is occupied at $V_{BG} = 0$ V when $t = 20$ s, while the first six subbands are occupied when $t = 2$ min. For the high electron density case, the electrons are squeezed to the surface due to the electron-electron interaction. This hollow circular density of states is similar to the model of which the nanowire is fully covered by the superconductor, namely full-shell nanowire.[52-53] In theory, the full-shell nanowires are supposed to enter the topological superconducting intervals when the applied parallel magnetic field corresponds to an odd multiple of the magnetic flux quantum, which implies a tiny magnetic field in the case of single flux quantum.[52-53] However, it suffers from the disadvantage of the absence of chemical potential tunability by electrostatic gate due to the screening of the full enclosing Al shell. Whereas, the partial-shell hybrid devices could mimic this kind of topological phase diagram while avoiding the disadvantage, when entering the regime with the ring-shape distribution of electron density. As a result, it might provide a more suitable platform to create and study MZMs.

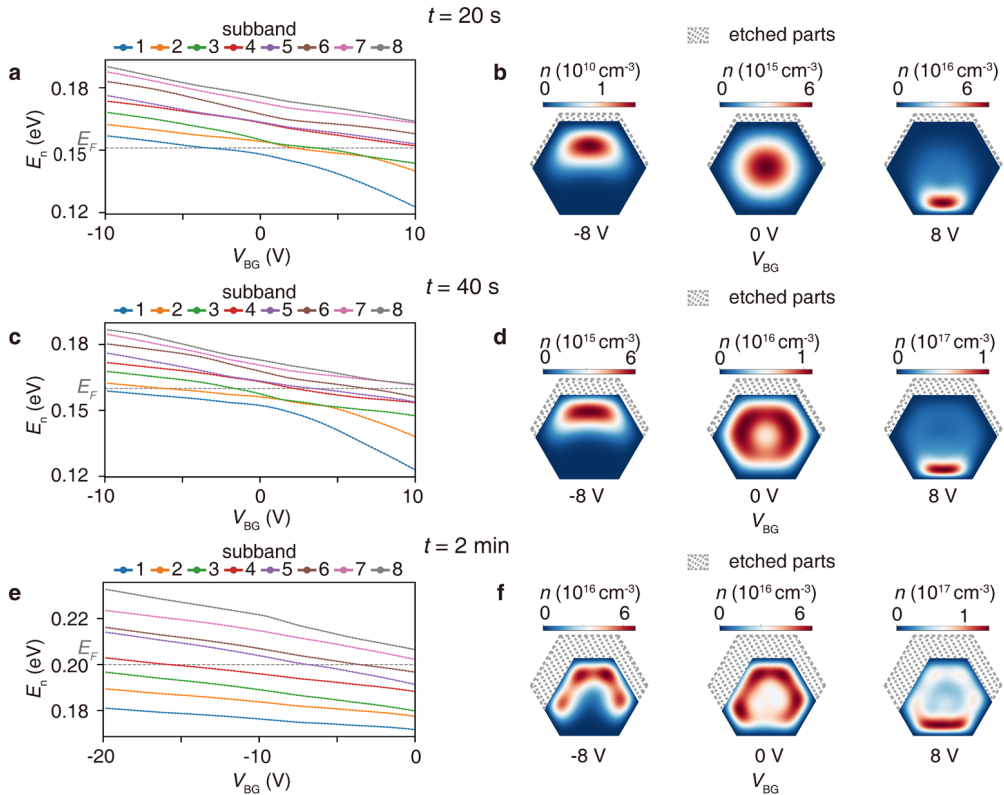

**Figure 3.** Calculated electronic structures and spatial distribution. a,b) The electronic structures and spatial distribution of the device etched for 20 s, calculated with Schrödinger-

Poisson method. The Fermi level is determined with the conductance at $V_{BG}$ = 0 V. The etched parts on the top and two upper side facets are indicated by gray lines but not with the real scale. c-f) The similar data of devices etched for 40 s and 2 min.

## 2.4. Modifying the Band Alignment at the Hetero-Interface by Argon Milling

Now let's discuss the detailed reasons for the modification of the wavefunction distribution by argon milling. As revealed in Figure 2d,f,h and Figure 3, the milling process can modify the electronic properties of the surface, and result in more negative back-gate voltage to pinch off the device, as well as the larger conductance at fixed gate voltage, with increasing milling time. The reason is either the screening effect due to the leads or greater band bending strength at the surface as modeled in Figure 3, and here we rule out the former again as follows.

The screening effect, which is actually the metal contacts reducing the effective field "felt" by the nanowire, leads to an increase in the absolute value of the threshold voltage no matter it is positive or negative and doesn't change its sign.[54] To check whether this fits our experimental result, we fabricated the JJ of different lengths ($l$) in series on the same nanowire with the same milling $t$ = 40 s. In this case, all the parameters, such as the details of the nanowire, would be strictly kept the same except for the length, which is equivalent to milling time. Because the under-cut structure (usually ~100 nm) of polymethyl methacrylate (PMMA) resist cause the nanowire regime between two leads exposed to the argon plasma during the milling process, as shown in **Figure** 4a. If the junction is short enough, the resist mask will form a bridge structure (indicated by the grey arrows), leaving more space under the resist to be etched. As such, we expect that shorter junction on the same nanowire to have more negative pinch-off $V_{BG}$, which is truly observed in Figure 4b. However, it should be noticeable that different from the negative pinch-off $V_{BG}$ in Figure 2d,f,h, that of the longer junction possess the absolutely positive values, and such change of sign contradicts with the screening effect. Moreover, once the pinch-off $V_{BG}$ shifts to be positive, the absolute value becomes smaller for short junction, which fit the overall trend but is contrary to the screening effect.

As a result, we attribute the reason to the more milling effect can bring about enhanced band bending of the accumulation layer at the surface, as illustrated in Figure 3. Such conclusion is further supported by the distribution of the conduction band edge relative to the Fermi level at $V_{BG}$ = 0 V in Figure 4c, which is calculated from self-consistently solving the Schrödinger-Poisson equation. This also explains the ring-shape electronic distribution which appears because of squeezing electrons to the surface due to Coulomb repulsion, when the electron density is high with long milling time.

For an ideal S-Sm junction, the band offset at the interface is determined by the difference of $\chi_{SM}$ and $W_M$, fitting Schottky-Mott rule. However, in reality it might deviates due to some non-ideal conditions, including the aforementioned Coulomb interaction inside in the case of high electron density and the surface states or adsorption of impurities on semiconductor.[55] Based on the clean interface confirmed by TEM and EDX, adsorption-induced band bending plays a minor role in the hetero-interfaces of our devices. The surface states, as well as the resulting position of the charge neutrality level relative to the Fermi level of the semiconductor, vary the charge density and electric field distribution, and finally have significant effect on the band alignment. Moreover, if their density of states is large, the Fermi level at the surface is "pinned" by the surface states near the charge neutrality level.[1] Consequently, we believe that the position of the charge neutrality level or the density of states of the surface states may change with the increase of the milling effect. This further results in a Fermi level pinning effect that tends to be above the conduction band bottom, manifested in enhanced band bending and stronger accumulation layers,[56] as shown in Figure 4d-f. It is worth noting that the strong pinning effect can shield the effect of metal contact, resulting in the band bending almost independent of the $W_M$.[1] Therefore, no matter whether the surface band bending is modulated through the strong electron interactions or the strong pinning effect brought about by adjusting the surface states, the subsequent aluminum film coverage will not change the fact that the band offset at the hetero-interface between the nanowire and superconductor varies with the milling time. In addition to metal contacts, high densities of surface states can also mitigate the effects of semiconductor contacts,[55] suggesting that our method may be applicable to conventional semiconductor devices.

In general, we can effectively change the band bending of Al-InSb by argon milling, which suffers from the natively negative band offset and weak coupling. A strong accumulation layer and strong coupling similar to in situ Al-InAs interface can be formed, which is essential for desirable high-quality hybrid devices.[25-30] Strong hybridization between the semiconductor and superconductor is considered as a critical factor in determining the quality of the induced gap. For different accumulation layers, the back-gate voltage expected to induce a hard gap varies significantly with the degree of hybridization. Also, the hybridization determines the renormalization of the Landé *g* factor and SOC strength, further controlling the topological phase diagram.[20-21, 31, 57]

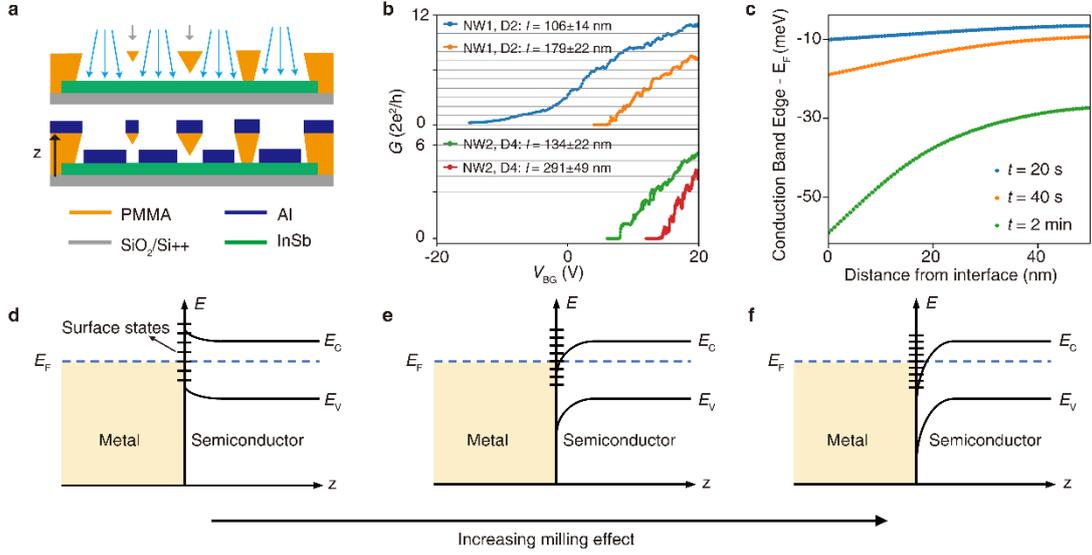

**Figure 4.** Band alignment at the hetero-interface. a) Schematic of the side view of the devices in the milling and deposition process. The grey arrows indicate the bridge structure formed by the resist mask in the short junction. b) Pinch-off curves for junctions with different widths in two nanowires (the horizontal line cuts from 2D maps at $V_{bias}$ = 0.47 mV). On the same nanowire, shorter junction has more negative pinch-off $V_{BG}$. c) Conduction band edge profile relative to the Fermi level in the semiconductor region for different milling time. d-f) Band diagram in the metal and semiconductor region for the different density of states of the surface states or the position of charge neutrality level. The short lines at the interface indicate the surface states.

## 2.5. High Tunable Nonlocal Signals

Here, we provide an example for constructing the advanced quantum devices on such system with strong coupling, which possesses the abovementioned combination of these high-quality components, i.e., a good induced gap and the strong band bending strength. An artificial Kitaev chian, which has been demonstrated to create Poor man's Majorana states, requires quantum dots with superconductor in-between to induce the coupling of ECT and CAR.[44, 58] In **Figure** 5a, we reveal the co-existence of both coherent processes on a three-terminal device (D4). By grounding the right electrode and applying a bias current $I_{bias}$ to the middle electrode, the relative position of the dc potential of these two electrodes can be tuned. And, the standard AC lock-in technique allows measurement of the nonlocal differential resistance $R_{NL} = dV_{NL}/dI \approx V_{NL}^{ac}/I^{ac}$ at different $I_{bias}$, where the nonlocal ac voltage $V_{NL}^{ac}$ is measured across the left junction. Because of the different directions of current brought by CAR and ECT for the nonlocal side, they result in different signs of $R_{NL}$.[59-61] We also apply the back-gate voltage to obtain $R_{NL}$ as a function of $V_{BG}$ and $I_{bias}$ (Figure 5b). The apparent CAR signal again proves the existence of a good induced gap.[17, 35] Meanwhile, the smooth transition between the CAR and ECT signals suggests the coexistence and high tunability of the two processes.

Specifically, the bias voltage where CAR or ECT dominates and distinguished by the sign of $R_{NL}$, switches between positive and negative $I_{bias}$ at different $V_{BG}$. The vertical line cuts at two typical $V_{BG}$ are shown in Figure 5c. Also, the switching points correspond to the Coulomb oscillation in the horizontal line cuts at $I_{bias} = 0$ nA. (the green curve shown in the bottom panel of Figure 5b). We attribute this feature to the formation of a quantum dot in the left junction and the energy-independent density of states in the right junction[16] (see Figure 5d,e, the thinner vertical lines on both sides of the right junction illustrate the low tunnel barriers where no quantum dot is formed). Despite superconducting electrodes on both sides, Coulomb charging energy can result in only a single resonance within the transport window, thereby suppressing the double ECT and double CAR processes involving multiple electrons.[62] When the resonance of the quantum dot is above the Fermi energy (dotted line), the ECT process will be more favorable in energy for positive $I_{bias}$ (Figure 5d), corresponding to negative $R_{NL}$ in the measurement. While increasing the $V_{BG}$, the resonance is tuned below the Fermi energy, and the CAR will be the dominant process in positive bias (Figure 5e), resulting a positive $R_{NL}$. Here, an asymmetric structure—QD-S-N is necessary and could be realized due to the shorter right junction than left junction. We mentioned above that a shorter junction needs a more negative gate voltage to pinch off, so the short junction is expected to be more difficult to reach the depletion region to form an unintended quantum dot at the same $V_{BG}$. To check this, with the conversion measurement configuration making the shorter junction as the nonlocal terminal, the signal associated with the quantum dot becomes vague or undetectable (Figure S5). The comparison further validates the conclusion above that the enhanced milling effect causes larger band bending. As such, it furnishes the essential for the investigation of Kitaev chains[44, 58] and Andreev molecules[63] which can also provide the basis for the extension of Andreev qubits.[8-10]

Furthermore, a minimal Kitaev chain has recently been achieved by strong coupling of ECT and CAR.[18] Notably, apart from a good induced gap and high tunability of the nonlocal signals, SOC is essential to ensure the presence of CAR between quantum dots with equal spins. When only spin-conserved tunnelling is allowed, a singlet CAR requires opposite spin states. SOC induces spin precession around the spin-orbit field axis in the hybrid segment, allowing a spin-up electron to acquire a finite spin-down component and pair with another spin-up electron to form a Cooper pair, known as a triplet CAR.[17] Although the hybridization with Al leads to the renormalization of InSb parameters which implies the lower effective $g$ factor and SOC strength,[20-21] the stronger electric fields due to the band bending at the interface will make the stronger SOC effects.[17] These desired characteristics agree with our devices and indicate that our versatile fabrication method is available for related experiments.

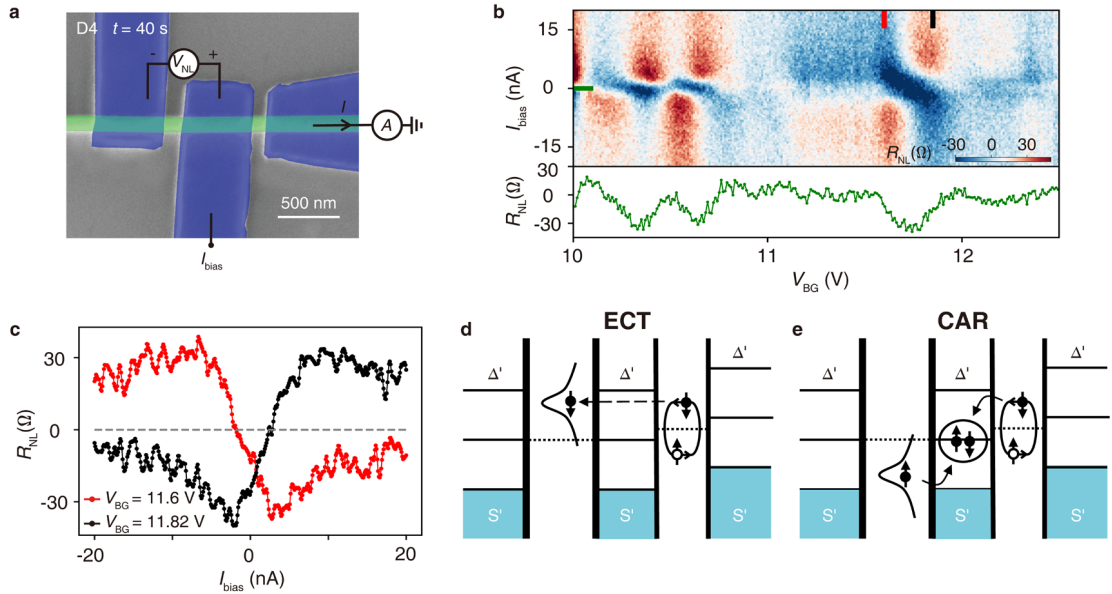

**Figure 5.** Nonlocal measurement of CAR and ECT (D4). a) The measurement configuration with bias current $I_{bias}$ and the nonlocal ac voltage $V_{NL}^{ac}$ across the left junction. b) Nonlocal differential resistance $R_{NL}$ as a function of $V_{BG}$ and $I_{bias}$. The green curve on the bottom panel shows the horizontal line cut from 2D maps at zero bias marked with the green bar. c) The vertical line cuts from panel (b) at two typical $V_{BG}$ marked with color bars. d,e) Schematic of ECT and CAR at positive $I_{bias}$ with different $V_{BG}$ corresponding to the two vertical line cuts in (c), respectively. The vertical solid lines indicate the tunnel barriers, and the lines' thickness represents the tunnel barriers' height. The horizontal dotted lines indicate the Fermi energy.

## 2.6. Hard and Large Induced Superconducting Gap in Tunnel Spectroscopy

For the tunnel spectroscopy measurements, we prepare the device (D5) in which the superconducting Al contacts the nanowire using the same fabrication method as the S-NW-S Josephson junctions described above (**Figure** 6a). Differently, the normal metal electrode titanium and gold (Ti/Au) is deposited on one side subsequently to form a normal metal-superconductor tunnel junction. Also, we make local bottom gates separated from the nanowire by boron nitride as the dielectric layer. Gate voltages on the local bottom gates, $V_{gate1}$ and $V_{gate2}$, are applied independently to control the chemical potential of different junctions. From the Josephson junction, we can obtain similar MAR traces as in Figure 2, proving that the high quality is maintained as the above devices (Figure S6a). Next, we perform the tunnel spectroscopy measurements of the normal metal-superconductor tunnel junction. In Figure 6b, as the bottom gate varies the chemical potential of the tunnel junction, the conductance as a function of $V_{bias}$ and $V_{gate2}$ showing the structure of the induced gap. The extracted value of the gap is about 0.17 meV, again similar to the gap of bulk Al. Meanwhile, the presence of ballistic transport in the tunnel junction is confirmed by the quantized above-gap conductance plateaus (the horizontal line cut from Figure 6b shown in Figure 6c). We also draw the vertical line cuts of two typical gate voltages in Figure 6d. In the open regime at high gate voltage, the subgap conductance is enhanced and higher than the above-gap conductance, results from Andreev enhancement at the strong coupling interface.[37, 39, 64-65] When the gate voltage is reduced so that the transmission is lowered to form a tunnel barrier, the subgap conductance exhibits significant suppression. However, compared to the above-gap conductance, the order of magnitude of suppression is not large enough to suggest a hard gap,[39-40] which we believe may originate from the lattice mismatch between the InSb and Al.[66-67] To improve the gap, a wetting layer[37, 39, 68] or hydrogen cleaning,[40-41] which might modulate the interface by high-temperature process, might help.

To verify the generic compatibility of the milling method and investigate better induced superconducting gap, we further fabricate the Pb-InSb hybrid devices (D6-8) with the optimized milling parameters. Pb has a superconducting gap of up to 1.4 meV and a higher critical temperature of 7.2 K,[69] which are almost seven times that of Al. However, due to the similar lattice mismatch between Pb and InSb, we add a Ti wetting layer after argon milling and before Pb film deposition (Figure 6e,f). The enlarged STEM-ABF image (contrast inverted) of the Pb film above the Ti wetting layer displays the ordered arrangement of the atoms over an extensive range, indicating the high quality of the Pb film (Figure 6g). The scanning electron microscope images of typical devices are shown in Figure 6h; and Figure

S6c. To check the induced superconducting gap, we measure the tunnel spectroscopy in the metal-superconductor tunnel junction shown in Figure 6i. A clean and enhanced superconducting gap up to seven times that of Al-based devices is observed. The vertical line cut at $V_{gate3} = 1$ V in log scale shows the ratio of above-gap and subgap conductance as high as two orders of magnitude, confirming the hardness of the gap (see the right panel of Figure 6i). Furthermore, the gap value (~1.4 meV) is close to the gap of bulk Pb, indicating the strong coupling between the nanowire and the superconductor. For the S-NW-S Josephson junction, Figure S6d shows the conductance as a function of $V_{bias}$ and $V_{BG}$, again supporting the hard and large superconducting gap (Figure S7). And the large induced gap is consistently observed across multiple devices (Figure S8). It is noteworthy that the large induced gap remained nearly unchanged for up to three months following device fabrication (Figure S9), demonstrating the long-term stability of the modified interface. This huge induced gap strongly extends the experimental parameter space for, e.g., the topological phase diagram, making the Pb-InSb hybrid device a promising platform for topological and Andreev qubits. Additionally, we fabricate Josephson junctions of varying lengths on the same InAs nanowire (Figure S10a,b). Conductance measurements reveal that shorter junctions exhibit higher conductance and more negative pinch-off back-gate voltages (Figure S10c,d). The similar band alignment modulation in the Pb-InAs system supports the potential application of our method to other hybrid quantum devices. Overall, our versatile method is compatible with different materials and facilitates the fabrication of hybrid devices containing various novel materials for exploring exotic physical phenomena.

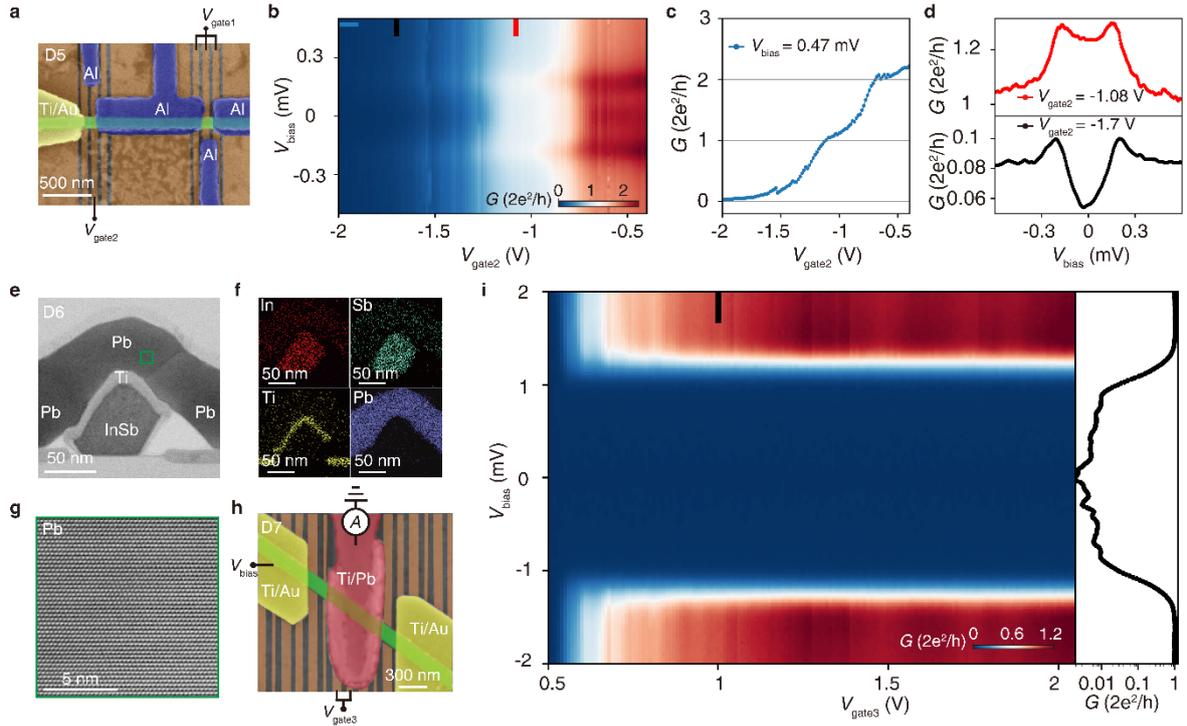

**Figure 6.** Tunneling spectroscopy of Al-InSb and Pb-InSb devices. a) False-color SEM image of Al-InSb device (D5). Ti/Au electrode (yellow) and Al electrodes (blue) contact the InSb nanowire (green). The local bottom gates (orange) are separated from the nanowire by boron nitride as the dielectric layer. b) Conductance as a function of $V_{bias}$ and $V_{gate2}$ for the normal metal-superconductor tunnel junction. c) Horizontal line cut from (b) shows the above-gap (blue, $V_{bias} = 0.47$ mV) conductance. d) Vertical line cuts from (b) in the tunneling regime (bottom panel) and open regime (top panel) show the induced superconducting gap and the Andreev enhancement, respectively. e) Cross-sectional STEM-ABF image of the device (D6). Due to the lattice mismatch between Pb and InSb, we add a Ti wetting layer (14nm) after argon milling and before Pb film (80nm) deposition. f) Corresponding EDX mapping for different elements. g) Enlarged STEM-ABF image (contrast inverted) of the Pb film at the location indicated by the green box in (e). To make all atoms visible in the image, we invert the contrast. The ordered arrangement of the Pb atoms persisted over an extensive range, indicating the high quality of the film. h) False-color SEM image of the device. Ti/Pb (red) and Ti/Au electrodes (yellow) contact the InSb nanowire (green), forming the Pb-InSb device. Labels indicate voltages applied on the local bottom gates ($V_{gate3}$). i) Conductance as a function of $V_{bias}$ and $V_{gate3}$ for the left tunnel junction of the Pb-InSb device (D7). The black curves on the right panel show conductance as a function of $V_{bias}$ in log scale.

## 3. Conclusion

In this paper, we use a generic method to achieve an atomic-resolved hetero-interface and high-quality hybrid devices. Gentle argon milling before depositing metal electrodes can remove the oxide layer and minimize the influence of the adsorption layer. Meanwhile, by well-controlled milling time, the generation of the accumulation layer makes the ohmic contact correspondingly formed at the interface, which can effectively avoid the common Schottky barrier between the metal and the semiconductor and fundamentally reduce the contact resistance. As a result, such Al-InSb hybrid devices display strong hybridization with a large induced gap, and ballistic transport with quantized plateaus. Importantly, by comparison with the theoretical model, our work demonstrates that this process can effectively modulate the band bending strength and the hybridized electron distribution, which ultimately determines the required performance of the devices, such as the CAR and ECT signified by nonlocal transport and the huge and hard induced superconducting gap by tunnel spectroscopy.

While some band-engineered heterostructures can currently be achieved by adjusting the chemical composition,[70] layer thickness,[71-72] and strain in the crystal structure,[73] these methods necessitate multiple material engineering and precise control over components. In particular, no experiments have been reported in semiconductor-superconductor hybrid nanowires, which are a potential platform for the realization of MZMs. However, tunable band alignment is crucial for exploring MZMs. Experimentally, we have observed the modulation of the band bending strength on both InSb and InAs. Our method facilitates straightforward modification of band alignment, introducing a new adjustable degree of freedom for investigating MZMs. Lastly, our approach can be integrated with state-of-the-art lithography fabrication techniques. Whether for the intricate architecture of qubits or Kitaev chains with increased quantum dots, our flexible fabrication process and the high-quality properties of our devices demonstrate significant potential for applications.

## 4. Experimental Section/Methods

*Device fabrication*: The InSb nanowires are grown by MOVPE[46] and transferred from the growth chip to the silicon substrates with micro-manipulator tip under the optical microscope. The silicon substrates are heavily doped and covered with 300 nm-thick $SiO_2$, which works as the dielectric layer of the back-gate. For the device with local bottom gates, the nanowires are transferred to pre-patterned Ti/Au gates. Boron nitride, as the dielectric layer, separates the gates from the nanowire. Prior to the transfer of both scenarios, we carry out the ozone cleaning on the substrates. The patterns of leads are defined with EBL. Before evaporation with metal, the nanowires are etched by argon plasma. The specific parameters for argon etching are beam voltage (260 V), beam current (10 mA), and air pressure (~$10^{-2}$ Pa), which are easily attainable in most argon milling instruments. The vacuum connection between the milling and evaporation chambers can effectively avoid oxidation or adsorption of impurities on the etched surface. Pb films tend to form discontinuous islands during deposition, which can affect the proximity superconductivity of devices. For the formation of a uniform superconducting layer, the temperature is kept around 130 K during the deposition process by means of liquid nitrogen cooling. Meanwhile the higher deposition rate (~0.5 nm/s) significantly improved the Pb film quality. Another issue to be aware of is that Pb film oxidises easily and can be etched by water. To avoid degradation of Pb film, the Pb is covered by 15 nm of $Al_2O_3$ without breaking the vacuum. Finally, the samples are naturally warmed to room temperature for more than 8h before unloading.

*Transport measurement*: The devices are measured at a base temperature of ~10 mK in an Oxford dry dilution refrigerator. Standard DC+AC lock-in technique allows measurement of the differential conductance and resistance. Typically, low frequency of ~13 Hz and AC excitation amplitude of ~10 μV or ~1 nA were used for lock-in measurement. The direction of the magnetic field was aligned with respect to the nanowire by detecting the gap value while rotating the magnetic field direction.

**Supporting Information**

Supporting Information is available from the Wiley Online Library or from the author.

**Acknowledgements**

We acknowledge fruitful discussions with Yu Liu and Chunxiao Liu. This work is supported by the National Natural Science Foundation of China (Grant Nos. 12174430, 92065203, 11527806, 12074417, 11874406, 11774405 and 12374459), the Strategic Priority Research Program B of the Chinese Academy of Sciences (Grants Nos. XDB33000000, DB28000000, and XDB07010100), the Beijing Nova Program (Grant No. Z211100002121144), the Synergetic Extreme Condition User Facility (SECUF), the National Key Research and Development Program of China (Grant Nos. 2016YFA0300601, 2017YFA0304700 and 2015CB921402), the NSFC for Young Scholars (Grant No. E2J1141), the Innovation Program for Quantum Science and Technology (Grant No. 2021ZD0302600), the program

"Excellence initiative - research university" for the AGH University of Krakow, the Beijing Natural Science Foundation (Z190010), the National Natural Science Foundation of China (52072400 and 52025025). D.P. acknowledges the support from the Youth Innovation Promotion Association, Chinese Academy of Sciences (Nos. 2017156 and Y2021043).

**Conflict of Interest**

The authors declare no competing interests.

**Author Contributions**

G.A.L., X.F.S., T.L., and G.Y. contributed equally to this work. J.S. conceived and designed the experiment. J.Y.S., D.G.Q., F.L., G.A.L., X.F.S. and A.-Q.W. fabricated devices with the help of G.Y. and Z.Y.Z., G.A.L., X.F.S., G.Y., J.Y.S., D.G.Q., F.L. and A.-Q.W. performed the transport measurements, supervised by B.B.T., P.L.L., Z.Z.L., G.T.L, F.M.Q., Z.W.D., Q.H.Z., L.L. and J.S., G.A.L., X.F.S., G.Y., M.P.N., P.W. and J.S. analyzed the data. M.R., G.B. and E.P.A.M.B. provided InSb nanowires. D.P. and J.H.Z. provided InAs nanowires. M.P.N. and P.W. fitted the MAR traces and calculated the electronic structures. T.L., L.G. and Q.H.Z. provided the TEM analysis. G.A.L., G.Y., M.P.N., P.W. and J.S. wrote the manuscript, and all authors contributed to the discussion of results and improvement of the manuscript.

# Supporting Information

**Title** Versatile Method of Engineering the Band Alignment and the Electron Wavefunction Hybridization of Hybrid Quantum Devices

*Guoan Li, Xiaofan Shi, Ting Lin, Guang Yang, Marco Rossi, Ghada Badawy, Zhiyuan Zhang, Jiayu Shi, Degui Qian, Fang Lu, Lin Gu, An-Qi Wang, Bingbing Tong, Peiling Li, Zhaozheng Lyu, Guangtong Liu, Fanming Qu, Ziwei Dou, Dong Pan, Jianhua Zhao, Qinghua Zhang, Erik P. A. M. Bakkers, Michał P. Nowak,\* Paweł Wójcik,\* Li Lu,\* and Jie Shen\**

## 1. Fitting of the Multiple Andreev Reflection (MAR) Traces

We numerically estimate the voltage bias-dependent conductance of the multimode nanowire Josephson junction as a sum of *M* single-mode contributions:[1-2]

$$G(V_{bias}) = \sum_{i=1}^{M} G_i(V_{bias}, T_i, \Delta') \qquad (S1)$$

where the fitting parameters are: Δ' — the superconducting gap of Al and $T_i$ — the transmission probabilities of the modes. We obtain Δ' and $T_i$ by minimizing the least-squares distance between the experimental and theoretical traces. We choose *M* such that we find at least one of the $T_i$ parameters remaining at zero within the considered $V_{BG}$ range.

## 2. Description of Schrödinger -Poisson Method

To obtain the electronic states of nanowire with different milling time we employ a standard envelope function approach in a single parabolic band approximation. Electron-electron interaction is included based the mean-field approximation by the standard Schrödinger-Poisson approach. If we assume the translational invariance along the growth axis z, the 3D Hamitonian reduces to 2D problem defined in the (x,y) plane

$$\left[-\frac{\hbar^2}{2m^*}\nabla^2_{2D} + E_c + V(x,y)\right]\psi_n(x,y) = E_n\psi_n(x,y) \quad (S2)$$

The eigenproblem (S2) is solved numerically on a triangular grid to avoid numerical artifacts coming from the discretization of the hexagonal-like area by a rectangular lattice. The hard wall boundary conditions are adopted which requires zeroing of the wave function at the boundaries of the sample whose geometry changes depending on the milling time (sizes of cross sections taken from the STEM images).

The eigenfunctions evaluated from Equation S2 is then used to determine the free electron densities which is given by

$$n_e(x,y) = 2\sum_n |\psi_n(x,y)|^2 \sqrt{\frac{\overline{m}^* k_b T}{2\pi\hbar^2}} \mathcal{F}_{-\frac{1}{2}}\left(\frac{-E_n + \mu}{k_b T}\right) \quad (S3)$$

where $\overline{m}^*$ is the effective electron mass along the nanowire axis, $k_B$ is the Boltzmann constant, $T$ is the temperature, $\mu$ is the Fermi level and $\mathcal{F}_k = \frac{1}{\Gamma(k+1)}\int_0^\infty \frac{t^k dt}{e^{t-x}+1}$ is the complete Fermi-Dirac integration of order $k$.

In the next iteration the Schrödinger Equation S2 is solved for the new potential including the electrostatic interaction of electrons at the mean free level. For this purpose, we solve the Poisson equation

$$\nabla^2_{2D} V(x,y) = -\frac{n_e(x,y)}{\epsilon_0 \epsilon} \quad (S4)$$

where $\epsilon$ is the dielectric constant. Equation S4 is solved by a box integration method on the triangular grid assuming Dirichlet boundary conditions which allows to include potentials applied to the nearby gates. The self-consistent Schrödinger -Poisson approach relies on the iteratively solution of Equation S2,4 until the convergence is reached.   Further details concerning the self-consistent method for hexagonal nanowires can be found in.[3] As mentioned in the main part of the paper the Fermi energy was determined in such a way to reproduce the conductance steps from the measurements. The calculations have been carried out for the parameters corresponding to the InSb material: energy gap $E_0 = 0.235$, $\epsilon = 16.8$ and $m^* = \overline{m}^* = 0.01359$.

In principle, the envelope function approximation assumes that the envelope of a forward-traveling wave varies slowly in space compared to the wavelength. This assumption is not always fulfilled in semiconductor nanostructures, especially when the size of the nanostructure (interfaces) reaches the value of several nanometers. In our case, however, the size of the smallest device is about 60 nm, compared to the Fermi length in InSb, which is of the order of 55 nm,[4] making the used approximation at the boundary of applicability. Note, however, that the Schrödinger-Poisson approach has been used over the years for many types of nanostructures, from semiconductor inversion and accumulation layers[5-6] through semiconductor heterojunctions and quantum wells[7-8] to semiconductor nanowires and quantum dots.[9-10] Although this approach has often been used beyond or at the border of its range of applicability, in many of these cases, this theory captures the electronic structure of low-dimensional semiconductor systems very well, and for this reason, it has become a standard tool simulations of semiconductor nanostructures.

**3. Calculation of the Mean Free Path**

To determine the mean free path le, we use the following equation:[11-12]

$$l_e = v_F \tau \quad (S5)$$

where $v_F$ is the Fermi velocity and $\tau$ is the momentum relaxation time. In the SP simulations, we use the parabolic band approximation so it means that the Fermi velocity $v_F = 1/(\hbar dE/dk)$ is given by $v_F = \hbar k_F/m^*$, where $k_F = \sqrt{2m^*(E_F - E_0)}/\hbar$ and $E_0$ is the bottom of the subband. We do this kind of calculations for the devices with different milling time from 20 s to 2 min (D1-3). Figures S3a-c present the Fermi velocity as a function of $V_{BG}$ calculated for each subband participated in the transport. Because electrons are injected in each of the occupied band, for a selected $V_{BG}$ we calculate the average velocity over the subbands (in black).

On the other hand, the momentum relaxation time $\tau$ and the field effect transistor (FET) mobility $\mu$ satisfy the following relationship:

$$\tau = \frac{\mu m^*}{q} \quad (S6)$$

where $q$ is the fundamental electronic charge.

In the Figure S3d-f, we measure the pinch-off single curves (not horizontal line cuts from 2D maps) above the gap (in red) for D1-3. To fit the mobility, we intercept the part of the high $V_{BG}$ at high $V_{bias}$ where the plateaus are averaged out due to many subbands being in the transport window between the Fermi levels of the source and drain, so the plateaus are less obvious. Specifically, we obtain the fitting curves (in black) by the following equation:[13]

$$G(V_B) = \left( R_s + \frac{L^2}{\mu C (V_B - V_{TH})} \right)^{-1} \quad (S7)$$

with the capacitance $C$ and the channel length $L$ where the fitting parameters are: $V_{TH}$ — the threshold voltage, $\mu$ — the field effect transistor mobility and $R_s$ — the interface resistances. The mobility estimated from the FET fit are $\mu_{20s} = 11954$ cm$^2$/Vs, $\mu_{40s} = 7843$ cm$^2$/Vs, and $\mu_{2min} = 6974$ cm$^2$/Vs, and the corresponding relaxation time are $\tau_{20s} = 0.092$ ps, $\tau_{40s} = 0.060$ ps, and $\tau_{2min} = 0.052$ ps, respectively. Finally, combining the above Fermi velocity and relaxation time, we find that the mean free path of the three devices is mostly between 100 nm to 200 nm [Figure S3g-i]. The characteristics larger than the length of junction ($l_{20s} = 109\pm12$ nm, $l_{40s} = 106\pm14$ nm, and $l_{2min} = 84\pm18$ nm) further demonstrate the ballistic transport of the devices.

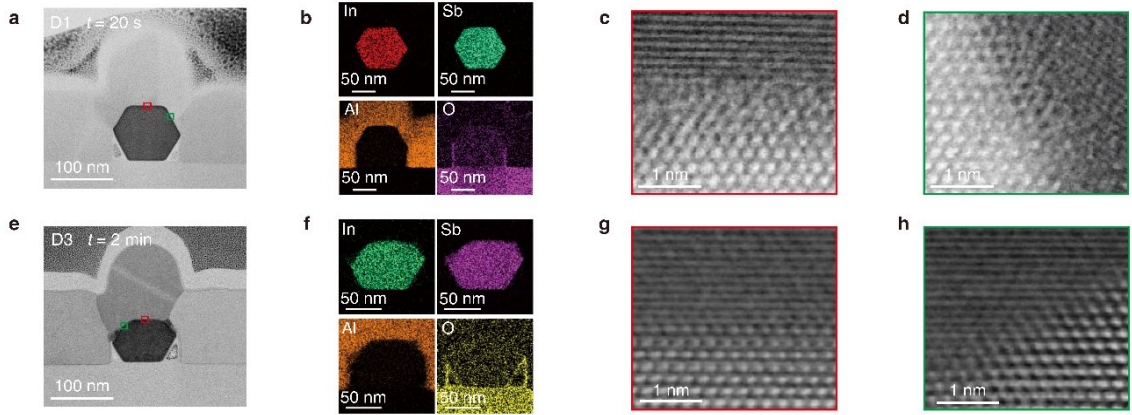

**Figure S1.** Material analysis of D1 and D3. a) Cross-sectional STEM-ABF image of D1 etched for 20 s. The specific thickness of the etched part at different milling time can be extracted from the image. b) Corresponding EDX mapping for different elements. The little distribution of oxygen at the hetero-interface between aluminum and nanowires demonstrates the nearly removal of the oxide layer. c) Enlarged STEM-ABF image (contrast inverted) of the top hetero-interface of Al-InSb at the location indicated by the red box in (a). d) Enlarged STEM-ABF image (contrast inverted) of the side hetero-interface of Al-InSb at the location indicated by the green box in (a). To make all atoms visible in the image, we invert the contrast. e-h) the similar data of D3 etched for 2 min. The hetero-interfaces on the top and sides remain of high quality even when the milling time is up to two minutes and part of the nanowires has been etched away.

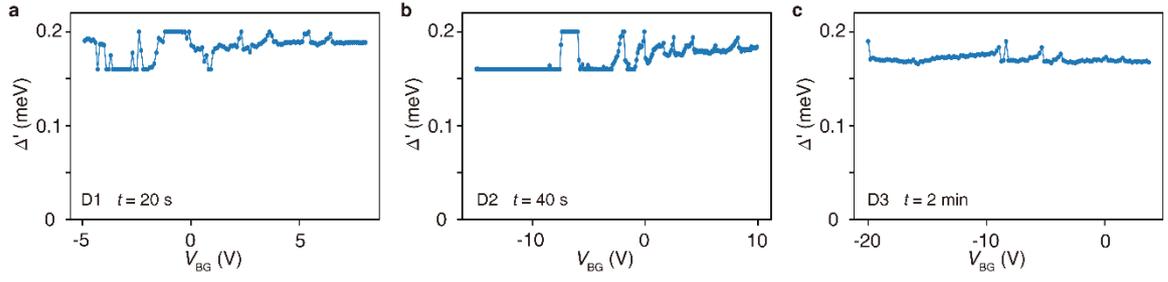

**Figure S2.** The value of the induced gap obtained from the MAR fit. a) The induced gap as a function of $V_{BG}$ of D1 etched for 20 s. b,c) The similar data of D2 and D3 etched for 40 s and 2 min. The induced gap is close to the gap of the parent Al, implying the existence of a transparent hetero-interface between the nanowire and the superconductor to guarantee a strong coupling between them.

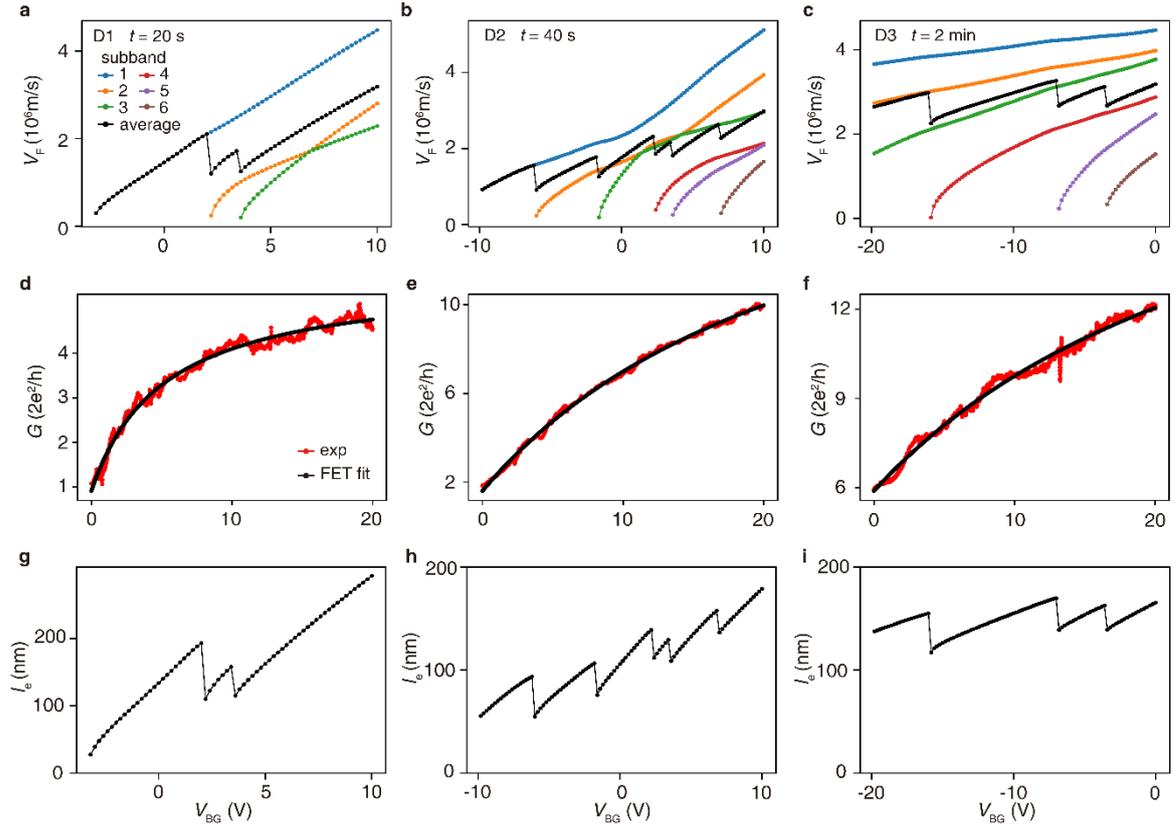

**Figure S3.** Calculation of the mean free path (D1-3). a-c) The Fermi velocity as a function of $V_{BG}$ for each subband participated in the transport calculated from SP simulations. The black curves indicate the average velocity over the subbands. d-f) Conductance as a function of $V_{BG}$ for these three devices above the gap (in red) and FET fit (in black). g-i) The mean free path as a function of $V_{BG}$ obtained by combining the Fermi velocity and momentum relaxation time.

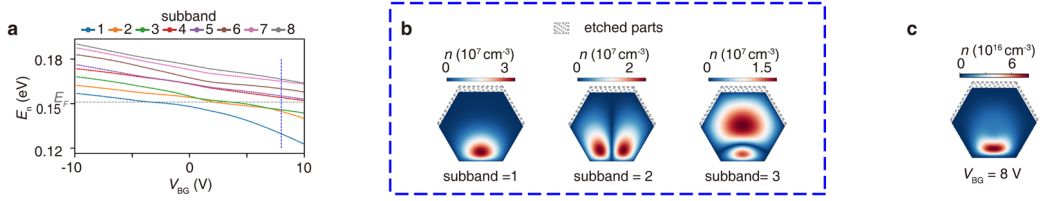

**Figure S4.** Correlation between the electronic structure and the spatial distribution. a) The electronic structures of the device etched for 20 s, similar to Figure 3a. The dotted line corresponds to $V_{BG} = 8$V. b) The wavefunction distribution of the individual electronic states below the Fermi energy at $V_{BG} = 8$V. The single wave functions are presented for ground subband = 1, and excited subband = 2,3 states for $k = 0$. Note however that, for $B = 0$ the system in translationally invariant which means that the wave functions do not change as we change $k$-vector. So those figures can be treated as results for $E_F$. c) The overall electronic spatial distribution acquired by integrating all wavefunctions below the Fermi energy.

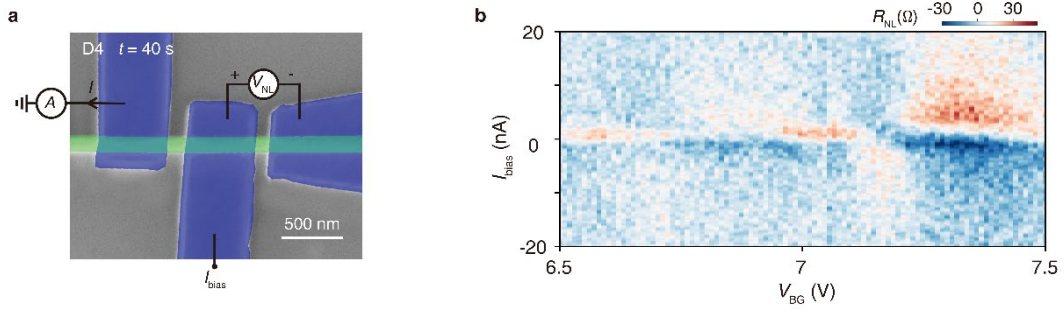

**Figure S5.** Nonlocal measurement of CAR and ECT different from the measurement configuration in the main text (D4). a) The measurement configuration with bias current $I_{\text{bias}}$ and the nonlocal ac voltage $V_{\text{NL}}^{\text{ac}}$ across the right junction. b) Nonlocal differential resistance $R_{\text{NL}}$ as a function of $V_{\text{BG}}$ and $I_{\text{bias}}$. The signal associated with quantum dots is not so clear compared to that in the main text. This should be related to the fact that quantum dot is not well defined at the short junction on the right, even at smaller $V_{\text{BG}}$.

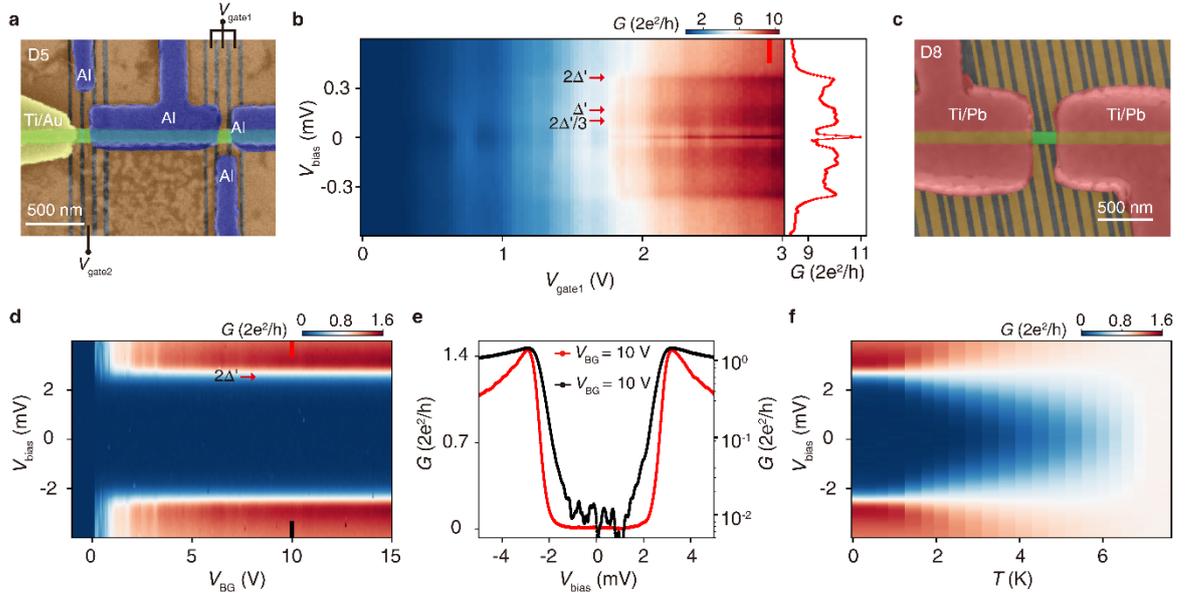

**Figure S6.** Transport measurement of Al-InSb and Pb-InSb Josephson junctions (D5, 8). a) False-color SEM image of Al-InSb device (D5). Ti/Au electrode (yellow) and Al electrodes (blue) contact the InSb nanowire (green). The local bottom gates (orange) are separated from the nanowire by boron nitride as the dielectric layer. b) Conductance as a function of $V_{bias}$ and $V_{gate1}$ for the Josephson junction. The red curve on the right panel shows the typical MAR curve as a function of $V_{bias}$, the vertical line cut from the 2D map, similar to the data in Figure 2. c) False-color SEM image of a typical Pb-InSb device (D8). Ti/Pb electrodes (red) contact the InSb nanowire (green), forming Pb-InSb Josephson junctions. d) Conductance as a function of $V_{bias}$ and $V_{BG}$ for a Pb-InSb Josephson junction. The typical 2Δ' (~2.8meV) structure shows up. e) The red (black) curves show conductance as a function of $V_{bias}$ with linear (log) Y axis, vertical line cuts from (d). The conductance suppression in the gap is about two orders of magnitudes, demonstrating a hard gap. f) Conductance as a function of $V_{bias}$ and temperature $T$ for the Pb-InSb device. The gap structure is fully suppressed at $T \sim$ 7 K, consistent with the critical temperature of Pb.

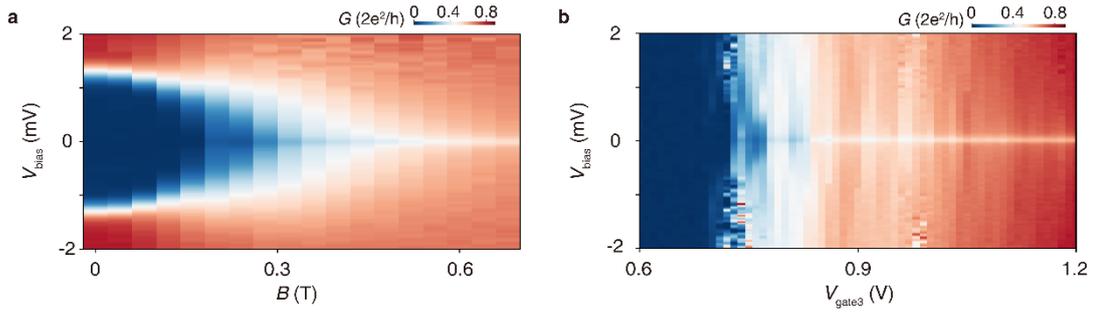

**Figure S7.** Transport measurement of Pb-InSb normal metal-superconductor tunnel junction (D7). a) Conductance as a function of $V_{bias}$ and magnetic field $B$ which is parallel to the nanowire for the normal metal-superconductor tunnel junction with $V_{gate3}$ = 1.5 V. b) Conductance as a function of $V_{bias}$ and $V_{gate3}$ for the normal metal-superconductor tunnel junction at $B$ = 1 T which is out of plane. The energy gap closes and disappears with applied magnetic field, indicating that the gap in the main text is an induced superconducting gap.

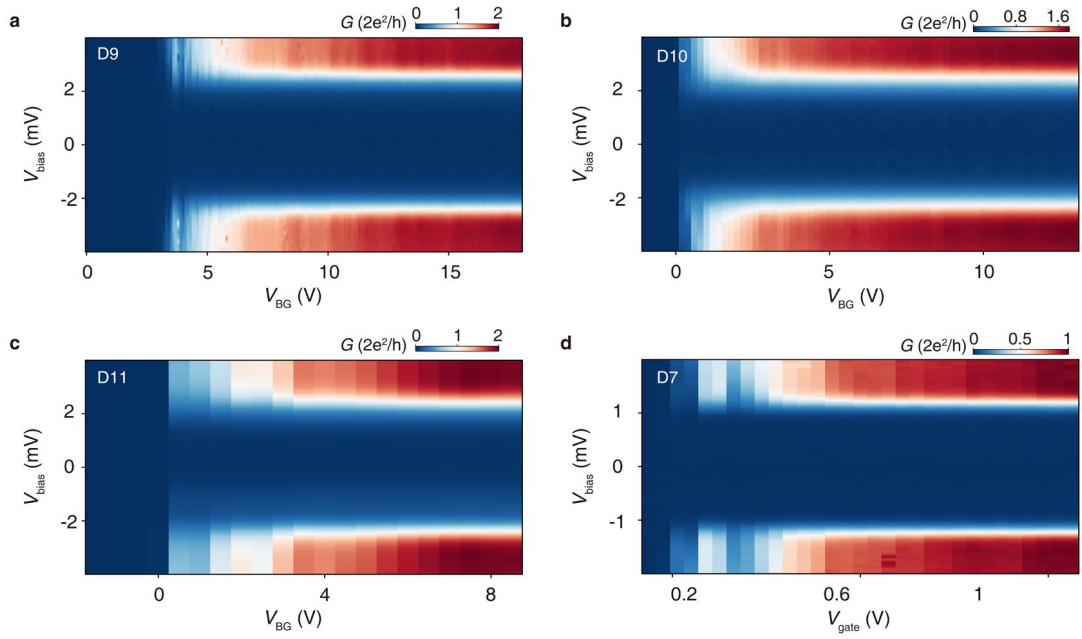

**Figure S8.** Tunneling spectroscopy of other devices. a-c) Conductance as a function of $V_{bias}$ and $V_{BG}$ for other Pb-InSb Josephson junctions (D9-11). d) Conductance as a function of $V_{bias}$ and $V_{gate}$ for the right tunnel junction of the Pb-InSb device (D7). The large induced gap is repeated in multiple devices.

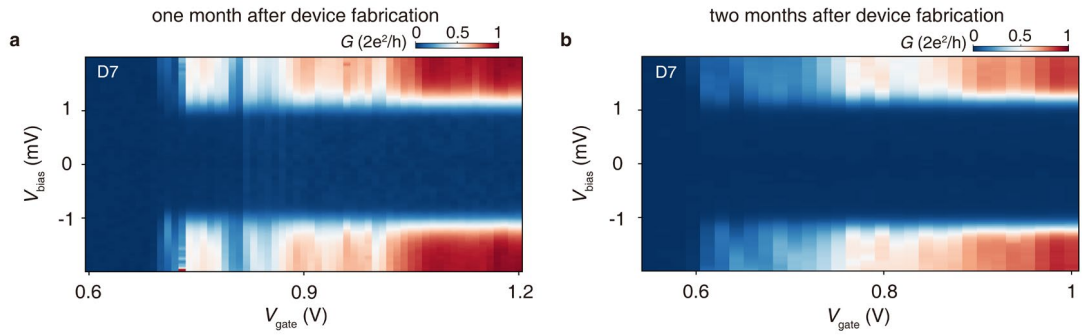

**Figure S9.** Long-term Stability of tunneling spectroscopy (D7). a) Conductance as a function of $V_{bias}$ and $V_{gate3}$ for the left tunnel junction of the Pb-InSb device (D7), measured one month after device fabrication. b) The similar data of D7 measured two months after device fabrication.

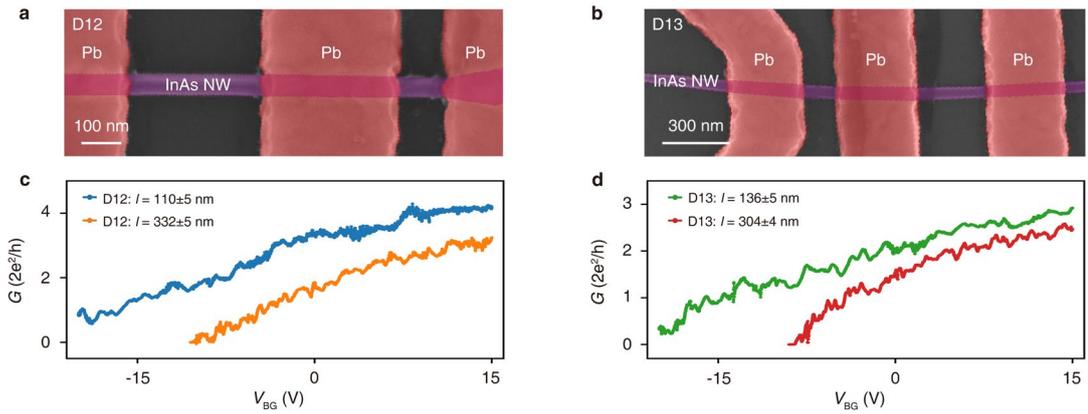

**Figure S10.** Transport measurement of Pb-InAs Josephson junctions (D12, 13). a) False-color SEM image of Pb-InAs device (D12). Three Pb electrodes (red) contact the InAs nanowire (purple). c) Pinch-off curves for junctions with different widths at $V_{bias}$ = 8 mV. On the same nanowire, shorter junction has more negative pinch-off $V_{BG}$. b,d) The similar data for D13.